\documentclass{article}

\PassOptionsToPackage{numbers, compress}{natbib}

\usepackage[final]{tackling_climate_workshop_style}

\usepackage[utf8]{inputenc} 
\usepackage[T1]{fontenc}    
\usepackage{hyperref}       
\usepackage{url}            
\usepackage{booktabs}       
\usepackage{amsfonts}       
\usepackage{nicefrac}       
\usepackage{microtype}      
\usepackage{graphicx}
\usepackage{gensymb}
\usepackage{csquotes}
\usepackage{arydshln}
\usepackage{tabularx}
\usepackage{rotating}

\title{Climate Impact Assessment Requires Weighting: \\ Introducing the \textit{Weighted Climate Dataset}}

\author{%
  Marco Gortan\thanks{These authors contributed equally.}\\
  Faculty of Business and Economics\\
  University of Basel \\
  \texttt{marco.gortan@unibas.ch} \\
  \And
  Lorenzo Testa$^*$ \\
  Department of Statistics and Data Science \\
  Carnegie Mellon University \\
  \texttt{ltesta@stat.cmu.edu} \\
  \AND
  Giorgio Fagiolo \\
  Institute of Economics and L'EMbeDS \\
  Sant'Anna School of Advanced Studies \\
  \texttt{g.fagiolo@santannapisa.it} \\
  \And
  Francesco Lamperti \\
  Institute of Economics and L'EMbeDS \\
  Sant'Anna School of Advanced Studies \\
  \texttt{f.lamperti@santannapisa.it} \\
}

\begin{document}

\maketitle

\begin{abstract}
High-resolution gridded climate data are readily available from multiple sources, yet climate research and decision-making increasingly require country and region-specific climate information weighted by socio-economic factors. Moreover, the current landscape of disparate data sources and inconsistent weighting methodologies exacerbates the reproducibility crisis and undermines scientific integrity. To address these issues, we have developed a globally comprehensive dataset at both country (GADM0) and region (GADM1) levels, encompassing various climate indicators (precipitation, temperature, SPEI, wind gust).
Our methodology involves weighting gridded climate data by population density, night-time light intensity, cropland area, and concurrent population count -- all proxies for socio-economic activity -- before aggregation. We process data from multiple sources, offering daily, monthly, and annual climate variables spanning from 1900 to 2023.
A unified framework streamlines our preprocessing steps, and rigorous validation against leading climate impact studies ensures data reliability. The resulting \textit{Weighted Climate Dataset} is publicly accessible through an online dashboard at \url{https://weightedclimatedata.streamlit.app/}.
\end{abstract}

\section{Background \& Motivation}

Climate change is a major driver of economic and social challenges \cite{dell2014we, carleton2016social}. Although research has documented its impacts on agriculture \cite{schlenker2009nonlinear}, economic production \cite{burke2015global, kalkuhl2020impact, kotz2022effect}, conflict \cite{abel2019climate}, income inequality \cite{palagi2022climate}, mortality \cite{carleton2022}, and energy consumption \cite{auffhammer2014measuring}, a critical data gap persists. Climate data are often available at high resolutions (daily or hourly) and gridded formats, while socio-economic data are typically collected annually and associated with administrative units like regions or countries.

To bridge this gap, researchers typically aggregate weather data to match the lower resolution of socio-economic data. However, this aggregation can be misleading if not done with care. Simply averaging weather data across a region can mask important spatial heterogeneity. For example, average summer temperatures in the Mojave Desert may be much higher than Los Angeles, but socio-economic activity is concentrated in Los Angeles. So, labor productivity in California is likely more affected by temperatures in Los Angeles than in the desert. Therefore, researchers are increasingly using spatially weighted climate data to account for the varying distribution of socio-economic activities within a region. Weighting schemes such as population density can be used to prioritize areas with higher socio-economic activity when averaging weather data. 

Despite the benefits, replicating studies using weighted climate data can be difficult. The specific weighting procedures used are often poorly documented or absent entirely in existing research.  This lack of a transparent, standardized, documented, and open access source for spatially weighted climate data can lead to biased results, as the choice of weighting methodology can significantly impact analysis, hindering accurate and robust estimation of climate impacts \cite{wei2022comparison}. This paper, by extending a previously published version \cite{gortan2024unified}, addresses this gap by introducing the \textit{Weighted Climate Dataset}, a unified source of data that preprocesses and weights gridded climate data. We also provide a user-friendly interface at \url{https://weightedclimatedata.streamlit.app/} in order to explore and download ready-to-use, spatially weighted climate variables at national and sub-national levels for the period 1900-2023. This promotes:
\begin{itemize}
    \item \textbf{Replicability:} Improved documentation and transparency in data processing;
    \item \textbf{Efficiency:} Saves researchers time and resources by providing pre-processed data;
    \item \textbf{Robustness:} Encourages testing the sensitivity of estimates to weighting choices.
\end{itemize}

By facilitating access to a standardized dataset, we aim to enhance the climate impact assessment community's work and encourage robust and replicable research. 

\section{The \textit{Weighted Climate Dataset}}

\begin{figure}
    \centering
    \includegraphics[width=\linewidth]{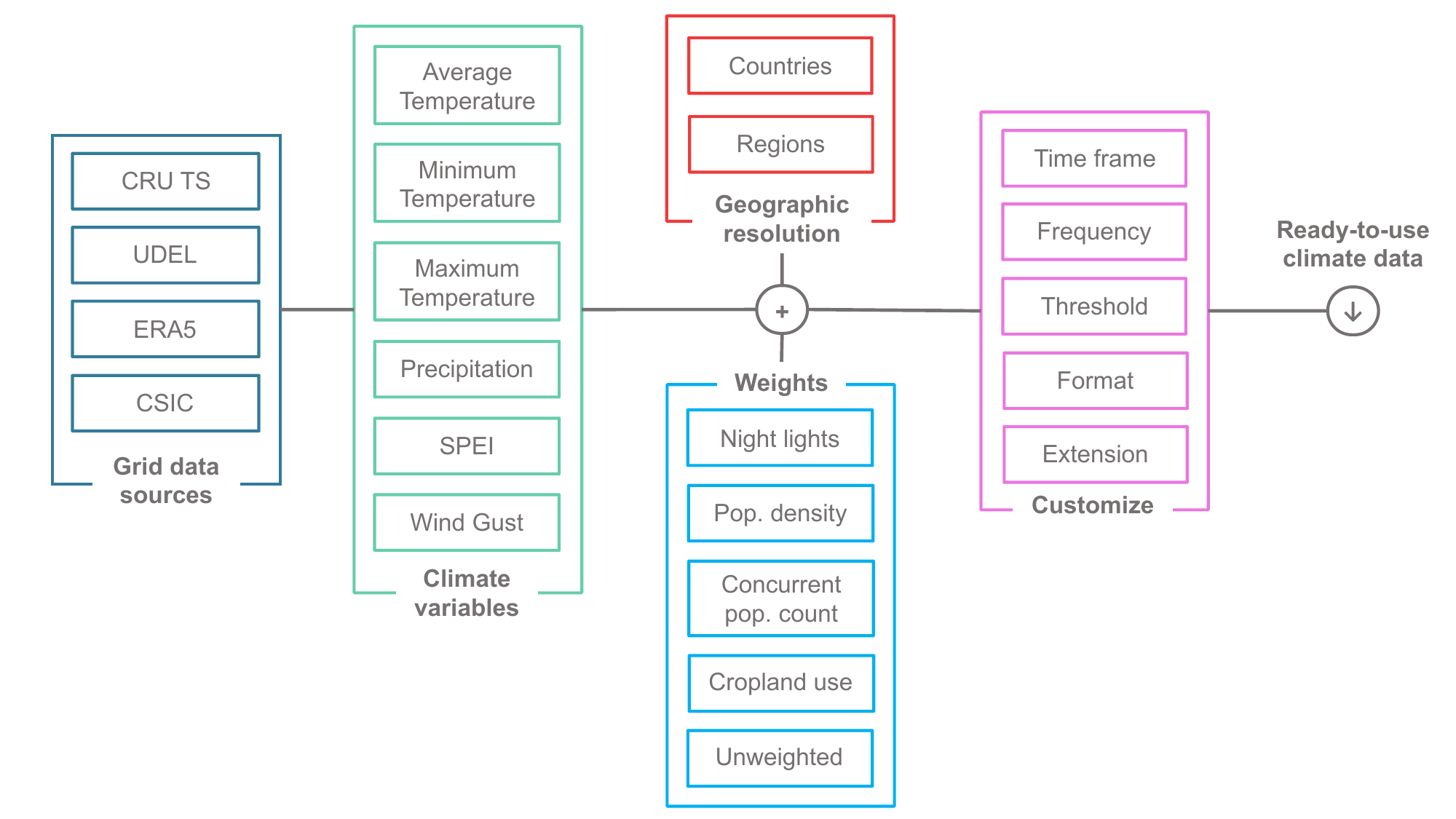}
    \caption{The \textit{Weighted Climate Dataset} workflow combines climate data, socio-economic indicators, and administrative boundaries to produce region-specific climate variables \textit{weighted} by socio-economic activity.}
    \label{fig:pipeline}
\end{figure}

Figure \ref{fig:pipeline} outlines the construction of our dataset. We integrate gridded climate data from multiple open-access sources with gridded socio-economic indicators and administrative boundaries. Our methodology enables users to customize the weighting of climate data by various socio-economic factors to suit their specific research needs.

\subsection{Sources}

\textbf{Climate Data.} We incorporate raw gridded climate data from four established datasets commonly employed in climate impact research: Climate Research Unit Time-Series \cite{harris2020version} (CRU TS v4.07), available from 1901 to 2022; Consejo Superior de Investigaciones Científicas \cite{vicente2010new} (CSIC v2.7), spanning 1901 to 2020; ECMWF Reanalysis v5 \cite{hersbach2020era5} (ERA5), covering 1940 to 2023; and University of Delaware \cite{willmott2000terrestrial} (UDEL v5.01), ranging from 1900 to 2017. CRU TS, UDEL, and CSIC offer monthly data at a spatial resolution of $0.5^{\circ}\times0.5^{\circ}$, whereas ERA5 provides data at a higher resolution of $0.25^{\circ}\times0.25^{\circ}$. 
All sources provide \textit{monthly} records of average temperature (in Celsius degrees, $\degree C$) and total precipitation (in millimeters, $mm$), with the exception of CSIC, which uniquely offers an additional monthly climate variable: the Standardized Precipitation-Evapotranspiration Index \cite{vicente2010multiscalar}, also known as SPEI (unit free). 
In addition to monthly data, ERA5 also provides records at the temporal resolution of \textit{hours}. We aggregate these hourly data into \textit{daily} values. Specifically, we calculate daily total precipitation (by taking the sum of hourly measurements), and average, minimum, and maximum temperature. Finally, ERA5 also provides 10-meter hourly wind speed measurements (measured in meters per second, $m/s$), which we aggregate into daily maximum wind gust.

\textbf{Socio-economic Data.} To capture the spatial distribution of economic and human activity, we incorporate five weighting schemes based on socio-economic variables. 
First, population density data are available from Columbia University's Gridded Population of the World v4 (GPWv4) \cite{doxsey2015taking}, and measured at $0.25^{\circ}$ and $0.5^{\circ}$ spatial resolutions. To accurately represent population size, we multiply population density by grid area.
Second, night-time light data \cite{li2020harmonized}, measured in digital number values (DN) and originally available at a 30 arc-second spatial resolution ($0.0083^{\circ}$), are incorporated after averaging to $0.25^{\circ}$ and $0.5^{\circ}$ resolutions. 
Third, cropland data are available from the History Database of the Global Environment (HYDE) \cite{klein2017anthropogenic}, version 3.2, and are measured in square kilometers. These data are recorded at a spatial resolution of 5 arc-minutes ($0.083^{\circ}$), and are incorporated after averaging to $0.25^{\circ}$ and $0.5^{\circ}$ resolutions.
Users can select population density, night-time light, or cropland as weighting factors, with base years of 2000, 2005, 2010, or 2015.
Fourth, climate data can be weighted solely by grid cell area, an option referred to as \textit{unweighted}.
Finally, for a dynamic approach, we provide a \textit{concurrent} weighting strategy where climate variables are weighted by population counts from the beginning of the respective decade. For instance, temperatures in 1907 are weighted using population data from 1900. Population counts are derived from HYDE v3.2 \cite{klein2017anthropogenic} for 1900-2010 and GPW for 2020.

\textbf{Administrative Boundaries.} We utilize two administrative levels from the Global Administrative Areas Database (GADM v4.1, released July 16, 2022) \cite{gadm}. The coarser GADM0 level corresponds to country boundaries, while the finer GADM1 level represents the largest subnational administrative units (e.g., states for the US).

A summary of the data sources and further information about our preprocessing pipelines can be found in the Supplementary Material.

\subsection{Aggregation}
We employ a general weighted aggregation scheme to calculate climate variables at the administrative unit level. The weighted value, $y_{i,t,w,T}$, for climate variable $y$ in geographic unit $i$ (at a specified GADM resolution) at time $t$ using weight $w$ measured in base year $T\in\{2000,2005,2010,2015\}$ is computed as
\begin{equation}
    \label{eq:aggregation}
    y_{i,t,w,T} = \frac{\sum_{j \in J_i} a_{j} f_{i,j} w_{j,T} x_{j,t}} {\sum_{j \in J_i} a_{j} f_{i,j} w_{j,T}}\,,
\end{equation}
where $J_i$ represents the set of grid cells intersecting unit $i$, $f_{i,j}$ is the fraction of grid $j$ within unit $i$, $a_{j}$ is the area of grid $j$, and $x_{j,t}$ is the raw gridded climate value. Typically, the base year $T$ is fixed, aligning with common practices in the literature \cite{burke2015global,auffhammer2018quantifying}. However, the \textit{concurrent} weighting scheme uses a dynamic base year, setting $T$ as the start of the decade containing $t$. In the \textit{unweighted} case, $w_{j,T}$ equals 1 for all $j$ and $T$. For cropland and concurrent weighting, the grid area term $a_{j}$ is set to 1. 

\begin{figure}
    \centering
    \includegraphics[width=\linewidth]{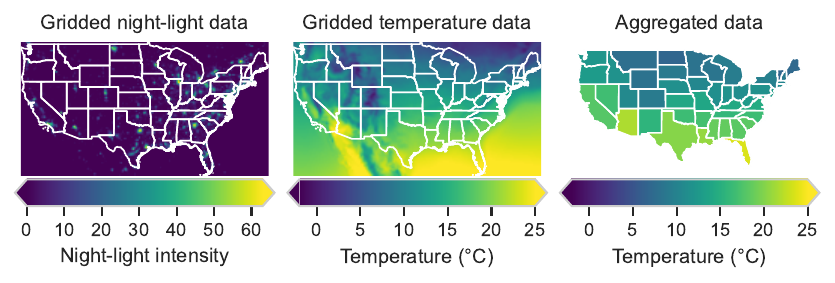}
    \caption{Climate data weighting example for the US in 2015. The left panel shows raw gridded night-time light data. The middle panel displays raw gridded average temperature. The right panel presents average temperature aggregated to the GADM1 level, \textit{weighted} by night-time light intensity.}
    \label{fig:US_eg}
\end{figure}

As an example of the aggregation strategy, Figure \ref{fig:US_eg} shows three panels. The left and center panels display raw gridded data for night-time light intensity in 2015, and ERA5 average annual temperatures in 2015 for the contiguous US, respectively. Night-time lights are chosen as the weighting variable in this example. Figure \ref{fig:US_eg}, right panel, displays the resulting aggregation at GADM1 resolution and illustrates the output that users can retrieve from our dataset. 

In the Supplementary Material, aiming to support the reliability and effectiveness of our approach, we validate the quality of our dataset by comparing it with those employed in two influential climate econometric exercises \cite{burke2015global, kotz2022effect}.

\subsection{Customization}

The \textit{Weighted Climate Dataset} dashboard provides several ways to customize the aggregated data. Users can select specific time intervals and choose between \textit{daily}, \textit{monthly}, or \textit{annual} frequencies. Notice that annual data are derived by aggregating daily or monthly values (e.g., summing monthly total precipitation). The SPEI is an exception, available only monthly due to its complex calculation.

To analyze extreme weather events, users can also define \textit{absolute}, \textit{relative}, and \textit{cumulative} thresholds. The dashboard, by exploiting daily data provided by ERA5, then calculates the frequency of days exceeding these thresholds for each geographic unit, month, or year. For relative thresholds, percentiles are calculated based on the historical distribution for each region or country. For cumulative thresholds, instead of computing the frequency, the dashboard provides the sum of the residual values that exceed the threshold. 

Data can be downloaded from the dashboard in two different formats: the \textit{wide} format has geographical units as keys and values of a climate variable in different years as attributes; the \textit{long} format has geographical units and years as keys, and the value of a climate variable as only attribute. Data can also be downloaded in three different extensions (\texttt{csv}, \texttt{json}, and \texttt{parquet}).

\section{Discussion}
The \textit{Weighted Climate Dataset}  offers a valuable resource for the climate impact assessment community, providing a standardized framework for evaluating and refining models and methodologies. By democratizing access to a comprehensive suite of datasets, we aim to propel research on the complex interplay between climate change and socio-economic dynamics. This resource allows researchers to further investigate the mechanisms through which climate variability and extremes influence economic growth, inequality, and other critical social factors. Ultimately, by enhancing our understanding of these relationships, the \textit{Weighted Climate Dataset} can contribute to more informed decision-making, effective policy development, and public engagement in addressing the challenges posed by climate change.

\begin{ack}
This work is partially supported by Italian Ministry of Research, PRIN 2022 project ``ECLIPTIC''. All analyses were performed on the computational resources of L'EMbeDS Department at Sant'Anna School of Advanced Studies. 
\end{ack}

\bibliographystyle{unsrt}
\bibliography{bib}

\appendix

\section{Additional information on sources}

\subsection{Climate data}

CRU TS utilizes raw meteorological data collected from a vast network of weather stations. Monthly deviations from average climatic conditions, known as anomalies, are calculated and subsequently interpolated using a method called Angular-Distance Weighting (ADW) \cite{harris2020version}. ADW is employed to account for the varying size of grid cells on the spherical Earth, specifically by considering the cosine of the latitude for each grid cell. This cosine value serves as a proxy for the change in grid cell area with respect to latitude. It is recognized that grid cells near the equator encompass larger areas compared to those closer to the poles.

CSIC leverages the CRU TS dataset to generate the Standardized Precipitation Evapotranspiration Index (SPEI), a drought index that comprehensively assesses both the severity and duration of drought conditions by incorporating information on precipitation and evapotranspiration. This index represents a standardized adaptation of the widely used Palmer Drought Severity Index (PDSI), which considers the combined influences of precipitation and temperature on water availability. Due to its multi-scalar nature, the SPEI can distinguish between different types of drought; for the present study, we concentrate on the one-month timescale to examine fluctuations in headwater levels.

ERA5 is a climate dataset that incorporates data collected by radiosondes, battery-powered instruments carried aloft by weather balloons to measure atmospheric parameters such as temperature, wind speed, and humidity. The data transmitted from these radiosondes to ground stations are integrated into the ERA5 dataset alongside observations from satellites and surface-based instruments through the utilization of numerical weather models, resulting in a comprehensive representation of the Earth's climate system \cite{hersbach2020era5}. As a unique \textit{reanalysis} product, ERA5 combines climate models with historical observations to deliver (i) consistent data over time and (ii) enhanced accuracy in grid areas without measurement stations.

Finally, UDEL provides gridded climate estimates primarily based on station records obtained from various publicly accessible sources (e.g., Global Historical Climatology Network dataset  \cite{peterson1997overview}, Global Historical Climatology Network Monthly dataset \cite{lawrimore2011overview}, Daily Global Historical Climatology Network archive \cite{menne2012overview}). The interpolation process is carried out using the Shepard spatial interpolation algorithm \cite{shepard1968two}, modified to accommodate the near-spherical shape of the Earth.

\subsection{Socio-economic data}

\begin{figure}
    \centering
    \includegraphics[width=\textwidth]{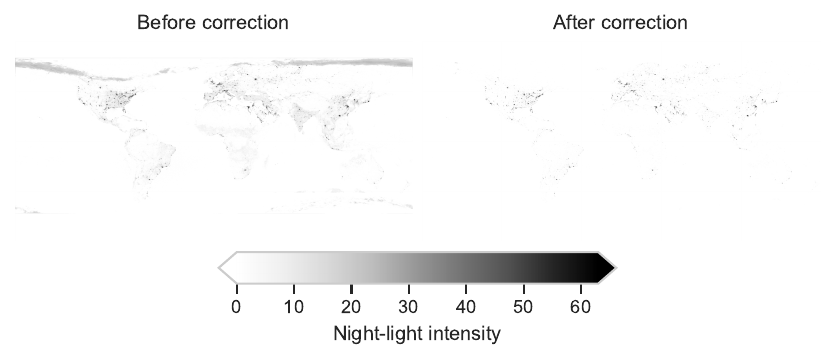}
    \caption{Correction of auroras and other noise sources in the night-time light data for the year 2015. The left plot shows night-time light data \textit{before} correction; the right plot shows the same data \textit{after} correction, which consists of setting to 0 the values in the grids whose digital number values are less than 30.}
    \label{fig:aurora}
\end{figure}

Night-time light and cropland data require preprocessing prior to analysis. Night-time light data consists of digital number (DN) values, a standardized measure of pixel brightness ranging from 0 to 63. Initially available at a high-resolution of 30 arc-seconds ($0.0083^{\circ}$), these data were aggregated to match the coarser resolutions of our gridded climate data ($0.25^{\circ}$ and $0.5^{\circ}$). This was accomplished by calculating the mean DN value for blocks of 900 (30$\times$30) and 3600 (60$\times$60) most upper-left cells in our coordinate system. We then iterate this procedure with the adjacent blocks of cells to obtain all gridded values of the night-light data for the coarser resolution. It is important to note that the harmonized VIIRS-DMSP dataset, particularly for the year 2015, contained noise from sources such as auroras and transient phenomena (e.g., boat lights, fires) as illustrated in Figure~\ref{fig:aurora} (left panel). To address this, following the approach of \cite{li2020harmonized}, DN values below 30 were set to zero prior to aggregation. The impact of this correction is visualized in Figure~\ref{fig:aurora} (right panel) for the year 2015.

Cropland data quantify the area, in square kilometers, dedicated to arable land and permanent crops within each grid cell. Originally recorded at a spatial resolution of 5 arc-minutes ($0.083^{\circ}$), these data were aligned with the coarser resolutions of our climate datasets through an aggregation process identical to that applied to the night-time light data.

A summary of the main characteristics of the original data sources is provided in Table~\ref{tab:sources}.

\begin{sidewaystable}
    \centering
    \caption{Summary of the main features of the employed data sources.}
    \label{tab:sources}
    \begin{tabularx}{\textwidth}{llXllll}
    \toprule
    \textbf{Source} & 
    \textbf{Reanalysis}
      & \textbf{Variables} & \textbf{Coverage period} & \textbf{Frequency} & \textbf{Resolution} & \textbf{Version} \\
    \midrule
    CRU TS \cite{harris2020version} & No & Average temperature, precipitation & 1901--2022 & Monthly & $0.5^{\circ}$ & 4.07 \\
    CSIC \cite{vicente2010new} & No & SPEI 1-month & 1901--2020 & Monthly & $0.5^{\circ}$ & 2.7\\
    ERA5\cite{hersbach2023era5} & Yes & Average temperature, minimum temperature, maximum temperature, precipitation, wind gust & 1940--2023 & Daily & $0.25^{\circ}$ & 5\\
    UDEL \cite{willmott2000terrestrial} & No & Average temperature, precipitation & 1900--2017 & Monthly & $0.5^{\circ}$ & 5.01 \\ 
    \midrule
    GPW
    \cite{doxsey2015taking} & & Population density & 2000, 2005, 2010, 2015 & Yearly & $0.25^{\circ}$ & 4 \\
    Li et al. (2020) \cite{li2020harmonized} & & Night-light intensity & 2000, 2005, 2010, 2015 & Yearly & $0.0083^{\circ}$ & 7\\ HYDE \cite{klein2017anthropogenic} & & Cropland & 2000, 2005, 2010, 2015 & Yearly & $0.083^{\circ}$ & 3.2 \\
    GPW
    \cite{doxsey2015taking} & & Population count & 2020 & Yearly & $0.25^{\circ}$ & 4 \\ HYDE \cite{klein2017anthropogenic} & & Population count & 1900-2010 & 10 years & $0.083^{\circ}$ & 3.2 \\
    \midrule
    GADM \cite{gadm} & & Administrative boundaries & & & & 4.1 \\
    \bottomrule
    \end{tabularx}
\end{sidewaystable}

\section{Further details on aggregation}

\begin{figure}[t]
    \centering
    \includegraphics[width=0.6\textwidth]{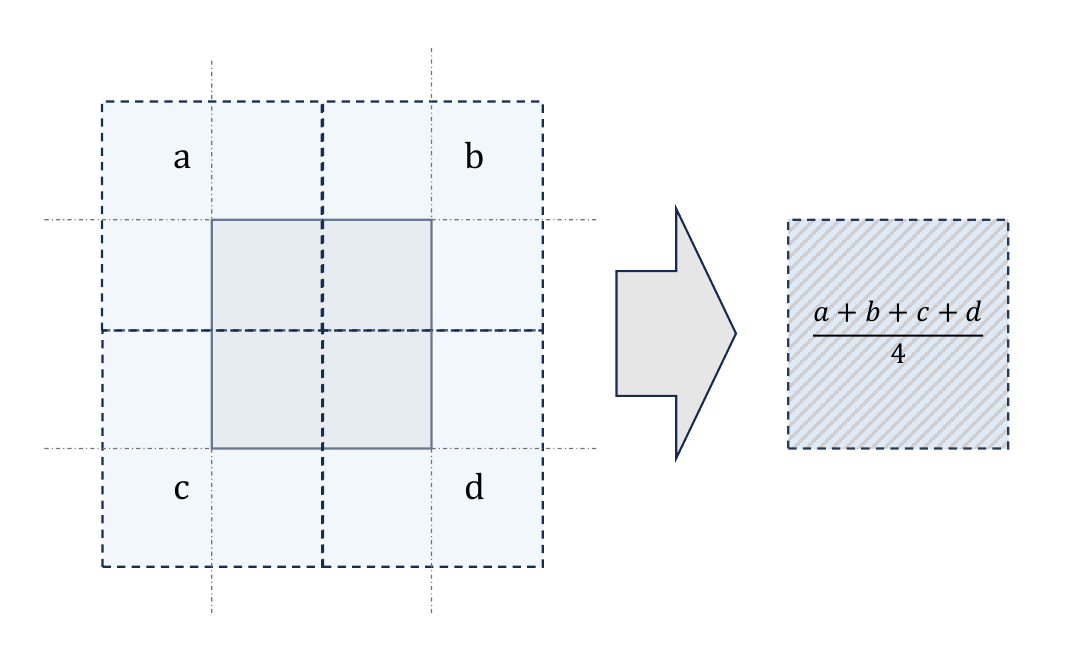}
    \caption{Stylized illustration of the resampling process applied to align weighting grids (population density, night-time light, cropland, and concurrent population count) with ERA5 climate variable grids. A slight spatial mismatch of $0.125^{\circ}$ exists between the grids in both longitude and latitude. To ensure accurate weighting, the weighting grids were resampled using a simple averaging method of overlapping grid cell values.}
    \label{fig:resample_ERA5}
\end{figure}

We observed inconsistencies in grid resolution across different datasets. ERA5 data are provided in a \texttt{NetCDF} file with a 721$\times$1440 grid, spanning from $180.125^\circ$W to $179.875^\circ$E and $90.125^\circ$S to $90.125^\circ$N, at a 15 arc-minute resolution. In contrast, socio-economic data at $0.25^\circ$ utilize a 720$\times$1440 grid, extending from $180^\circ$W to $180^\circ$E and $90^\circ$S to $90^\circ$N. To align ERA5 data with socio-economic variables for weighting and analysis, we employed bilinear interpolation to resample the ERA5 weight grids. This process is visually illustrated in Figure \ref{fig:resample_ERA5}, which compares the grid structures of the two data sources. This resampling procedure was applied consistently when combining ERA5 climate variables with population density, night-time light, cropland, and concurrent population count data.

Both climate and socioeconomic datasets exhibited sporadic missing values. To maintain data integrity, we adopted a conservative approach. When accurate climate variable weighting was hindered due to factors such as zero-valued weights or missing climate data within specific geographic units, we opted to retain the missing values (NAs) rather than imputing estimates.

\section{Validation}

This section validates our dataset against those used in influential climate econometric studies \cite{kotz2022effect, burke2015global}. By comparing our weighting procedures to theirs, we assess the reliability and effectiveness of our approach.

To ensure a rigorous validation, we replicated the exact data sources used in these studies, including older versions of climate and economic data. This enables us to verify the accuracy and robustness of our data processing pipeline and to objectively evaluate the quality and consistency of our estimates. \cite{burke2015global} utilized UDEL v3.01 for precipitation and temperature data, and GPW v3 at $0.50^{\circ}$ gridded population data for the year 2000. While the specific source and version of administrative boundaries are unspecified, their shapefiles are publicly accessible. In contrast, \cite{kotz2022effect} employed  $0.25^{\circ}$ gridded ERA5 precipitation and temperature data without applying any weighting, using GADM1 v3.6 for spatial aggregation in their primary analysis.

\begin{figure}
    \centering
    \includegraphics[width=\textwidth]{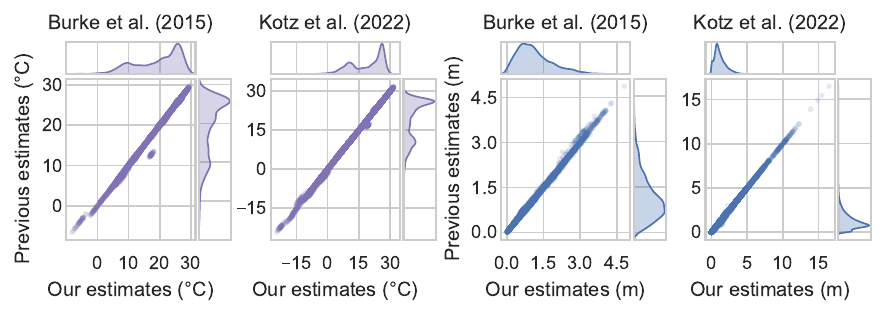}
    \caption{Comparison of weighted and/or aggregated average yearly temperature (degrees Celsius) and annual total precipitation (meters) variables in our datasets against data used in \cite{burke2015global} (countries) and \cite{kotz2022effect} (sub-national regions). Data points falling on the main diagonal indicate strong agreement between the estimates.}
    \label{fig:cmp}
\end{figure}

Figure \ref{fig:cmp} presents scatterplots comparing our average temperature and rainfall estimates with those from the original studies by \cite{burke2015global} and \cite{kotz2022effect} (SPEI, wind gust, and other summary statistics of temperature data were not used in either study). Data points aligned along the main diagonal of these plots indicate strong agreement between the estimates.
Our analysis reveals a high degree of concordance between our estimates and the weighted or aggregated data used by previous researchers across all study years. This supports the quality and reliability of our dataset and methodologies. However, minor discrepancies were observed. The first panel in Figure \ref{fig:cmp} highlights two key differences between our estimates and those of \cite{burke2015global}. Lower-left quadrant points represent negative temperatures in Greenland, where our estimates are exceeded by the ones of \cite{burke2015global}. Conversely, Bhutan exhibits slightly higher estimates in our data compared to \cite{burke2015global}. These discrepancies primarily stem from the weighting scheme, particularly the concentrated population density in specific regions of Greenland and Bhutan.

\end{document}